\documentclass{ws-ijgmmp}

\usepackage{xcolor}
\usepackage{booktabs}
\usepackage{multirow}
\usepackage[verbose,hypertexnames=false]{hyperref}
\hypersetup{colorlinks=false,allbordercolors=blue,pdfborderstyle={/S/U/W 1}}

\begin{document}

\markboth{D. Revanth Kumar, Santosh Kumar Yadav}
{Precision Constraints on New Dark Energy Parametrization from DESI BAO DR2}

%
\catchline{}{}{}{}{}
%

\title{Precision Constraints on New Dark Energy Parametrization from DESI BAO DR2}

\author{D. Revanth Kumar}

\address{Department of Mathematics, School of CS \& AI, SR University,\, \\ Warangal-506371, India.\\
\email{2406c9m003@sru.edu.in} }

\author{Santosh Kumar Yadav}

\address{Department of Mathematics, School of CS \& AI, SR University,\\
Warangal-506371, India.\\
sky91bbaulko@gmail.com }

\maketitle


\begin{abstract}
In this article, we investigate a new parametrization of the dark energy equation of state (EoS) with a single parameter for a barotropic fluid that deviates from the standard $\Lambda$CDM cosmology. We derive observational constraints on the model parameters using recent datasets including Observational Hubble Data (OHD), Pantheon+SH0ES (PPS), and Dark Energy Spectroscopic Instrument Baryon Acoustic Oscillations Data Release 2 (DESI BAO DR2). We constrain the best fit value of the parameter as, $\alpha =0.239 \pm 0.07$ at 68\% CL from joint analysis, which is non-null and suggests deviations from the cosmological constant. The model accommodates varying values of Hubble constant from different datasets and joint analysis yields $H_0 = 68.40 \pm 0.23$ $\mathrm{Km\,s^ {-1} Mpc^{-1}}$ at 68\% CL. We examine the physical behavior of the model by analyzing the deceleration parameter, the age of the universe, and the Om(z) diagnose. The deceleration parameter confirms a smooth transition from the past deceleration phase to the present cosmic acceleration as well as and also a second future transition back to deceleration, when PPS data is employed. 
\end{abstract}



\section{Introduction}	\label{sec:intro} 
Following the late 1990s discovery of cosmic accelerated expansion~\cite{Riess:1998cb,SupernovaCosmologyProject:1998vns}, several theoretical models have been proposed/investigated to explain this unexpected discovery of the universe. From these models, the standard model of cosmology commonly known as Lambda Cold Dark Matter ($\Lambda$CDM) is widely regarded by the research community as the simplest and most broadly accepted mathematical framework~\cite{Planck:2018vyg}. This $\Lambda$CDM paradigm consists
of two dominant components, namely, a non-luminous dark matter fluid and dark energy (DE) which together make up
around 95\% of the total energy budget of the universe. The DM fluid plays a substantial role in forming the universe's structure while the DE is assumed to be accountable for the current accelerating phase of our universe. Within the $\Lambda$CDM framework, DM is treated
as non-relativistic cold matter, referred to as CDM while DE is non-dynamic as a cosmological constant $\Lambda$, lacking any solid physical basis. The cosmological constant for DE represents vacuum energy with a fixed equation of state (EoS) $w_{\Lambda} = -1$. This $\Lambda$CDM paradigm offers an excellent fit to the cosmological data obtained from various investigations and a broad range of scales~\cite{Planck:2018vyg,DES:2018paw, Blomqvist:2019rah,DES:2021wwk, Riechers:2022tcj}. Nevertheless, it faces some serious challenges from both the theoretical and observational fronts. On one hand, $\Lambda$CDM cosmology suffers from initial challenges such as the cosmological constant problem~\cite{Weinberg:1988cp,Carroll:1991mt} and cosmic coincidence problem~\cite{Zlatev:1998tr}. On the other hand, with the consistent improvements and advancements in observational data, some key parameters show discrepancies when estimated from early universe measurements under the $\Lambda$CDM framework, such as those from Planck-CMB and from other astronomical probes. The most well-known discrepancy is the value of the Hubble constant $H_0$, commonly known as \texttt{Hubble tension}. The Planck CMB data prefer $H_0 = 67.36 \pm 0.54\,\mathrm{Km\,s^ {-1} Mpc^{-1}}$~\cite{Planck:2018vyg} within flat $\Lambda$CDM, while the SH0ES distance-ladder measurements yield $H_0 = 73.04 \pm 1.04\,\mathrm{Km\,s^ {-1} Mpc^{-1}}$~\cite{Riess:2021jrx}, with the latest SH0ES analysis reporting $H_0 = 73.17 \pm 0.86\,\mathrm{Km\,s^ {-1} Mpc^{-1}}$~\cite{Breuval:2024lsv}. The discrepancy is further amplified when comparing the recent Local Distance Network (H0DN) consensus value,
$H_0 = 73.50 \pm 0.81\,\mathrm{Km\,s^ {-1} Mpc^{-1}}$~\cite{H0DN:2025lyy},
with the combined Planck+ACT+SPT (SPA) constraint,
$H_0 = 67.24 \pm 0.35\,\mathrm{Km\,s^ {-1} Mpc^{-1}}$~\cite{SPT-3G:2025bzu},
corresponding to a tension exceeding $7\sigma$. The recent SPT-3G data alone yield $H_0 = 66.66 \pm 0.60\,\mathrm{Km\,s^ {-1} Mpc^{-1}}$ within $\Lambda$CDM, already $6.2\sigma$ below the SH0ES value~\cite{SPT-3G:2025bzu}. Taken together, these early–late universe discrepancies strongly suggest either unaccounted-for systematics or the need for new physics beyond the minimal $\Lambda$CDM model.

Many alternative approaches are proposed in the literature to address the issues of the $\Lambda$CDM model. These approaches can be classified into two broad categories: one mainly covers the modified gravity models~\cite{Copeland:2006wr,Clifton:2011jh,Koyama:2015vza,nojiri2017modified,koussour2023constraining,duchaniya2024cosmological} and the other is DE models in the context of Einstein’s general relativity~\cite{DeFelice:2010aj, Scherer:2025esj, Giare:2024gpk, deCruzPerez:2024shj, Yadav:2019jio, nagpal2025late}. Focusing on the second approach, the simplest modification of a $\Lambda$CDM cosmology is introducing a dynamical DE, where the DE EoS can be taken as a constant with a value different from $-1$ or a time-dependent function. Since there is no underlying fundamental principle to derive the DE EoS, several parametrized forms of the DE EoS have been introduced in the scientific literature. These parametrizations include one-parameter models~\cite{gong2005probing,yang2019observational}, two-parameter models such as the Chevallier-Polarski-Linder parametrization~\cite{chevallier2001accelerating, linder2003exploring}, exponential parametrization~\cite{Dimakis:2016mip, Pan:2019brc}, logarithmic parametrization~\cite{efstathiou1999constraining}, linear parametrization~\cite{cooray1999gravitational,
astier2001can,weller2002future}, Jassal-Bagla-Padmanabhan parametrization~\cite{Jassal:2005qc}, and Barboza-Alcaniz parametrization~\cite{barboza2008parametric} and many more in different context and perspectives. For a comprehensive list of parametrizations of DE EoS and other related details, we refer the readers to~\cite{Escamilla:2024fzq, nagpal2025late, Pacif:2020hai, abdalla2022cosmology}.

Motivated by the above mentioned parametrizations, we have considered a new parametrization of DE EoS from~\cite{singh2024new}. At high redshifts, this parametrization approaches a cosmological constant and hence retrieves the $\Lambda$CDM paradigm for the parameter value, $\alpha=0$. We have investigated the model by using more advanced and recent datasets, including Observational Hubble Data (OHD), Pantheon+SH0ES (PPS), and Dark Energy Spectroscopic Instrument Baryon Acoustic Oscillations Data Release 2 (DESI BAO DR2). In our analysis, we consider broader prior ranges for the model parameters as compared to the priors taken in~\cite{singh2024new}. The broad priors allow the MCMC algorithm to explore the full parameter space, capture possible degeneracies, and converge reliably toward the high-likelihood regions preferred by the data. Consequently, the final constraints reflect the true constraining power of the datasets, providing unbiased and physically meaningful estimates of the model parameters.
The main objective of this work is to constrain the free parameter of DE  EoS in light of the data sets used, to look at the possibility of the dynamical nature of DE, and to estimate the Hubble constant in this new parametrization of DE EoS with DESI BAO DR2. The physical properties of the model are investigated and discussed in detail to analyze the departure of the derived model from the standard $\Lambda$CDM cosmology. 
 
The work is structured as follows: In Section \ref{model_equations}, we present the governing model equations for the new parametrization. In Section \ref{model_data}, we mention observational data and the methodology used in the present work. Section \ref{results} presents the results and discussion. In Section \ref{model_properties}, we investigate the physical behavior of the model including the evolution of the deceleration parameter, age of the universe, and Om diagnose test. Finally, Section \ref{model_conclusion} presents the main findings of our investigation.

\section{Governing Model Equations} \label{model_equations}
We assume that the geometry of the universe is most accurately described by a spatially flat, homogeneous, and isotropic  Friedmann-Lema\^{i}tre-Robertson-Walker (FLRW) universe whose line element is given by (we adopted the speed of light $c=1$)
\begin{equation}
    ds^2 = -dt^2 + a^2(t)\bigg[dr^2 + r^2 (d\theta^2 + \sin^2 \theta \, d\phi^2)\bigg],
\end{equation}
where $a(t)$ characterizes the universe scale factor (henceforth abbreviated as $``a"$). The evolution of $a$ with redshift $z$ follows the relation: $a(t) = a_0/(1+z)$, $a_0$ denotes the scale factor at the present epoch. We fixed $a_0=1$ in the forthcoming text. From the above metric, the large-scale dynamics of the Universe can be derived using the fundamental equations, so-called Friedmann equations are given by
\begin{align}\label{model}
3 H^2 &=  8 \pi G (\rho_{\rm r} + \rho_{\rm m} +\rho_{\rm de}),\\ 
2 \frac{dH}{dt} + 3 H^2 &= - 8 \pi G (p_{\rm r} + p_{\rm m} + p_{\rm de}),
\end{align}
where $H$ refers to the Hubble parameter, which encapsulates the time rate of change of the cosmic scale factor, given by $H = \frac{da/dt}{a}$. The Newtonian constant is $G$, as usual. The density and pressure of radiation, matter, and DE are denoted by $\rho_i$ and $p_i$  where $ i \in \{ \rm r, \rm m, \rm de \}$ respectively. Hereafter, we have neglected the radiation component as its effects are negligible at late time and has no significant impact on the evolution equations.

In the absence of any interaction beyond gravity between the above-mentioned species, the conservation equation (commonly referred to as the fluid equation) for the $i^{\rm th}$ component reads as
\begin{align} \label{continiuty}
    \dot{\rho}_i + 3 \frac{\dot{a}}{a} (1+w_i)\rho_i = 0,
\end{align}
where  $w_i$ is EoS of the $i^{ \rm th}$ specie with relation, $p_i =w_i \rho_i$ for any barotropic fluid. We assume that the matter sector of the universe is made of pressureless matter (baryons + CDM) which leads to $w_{\rm m}=0$. and a DE fluid for which $w_{\rm de}(a)$ is dynamical. From eq.~\ref{continiuty}, one can easily derive the evolution of radiation and matter. Finally, for a dynamical DE with EoS $w_{\rm de}(a)$ , its density evolution can be obtained by following
\begin{align} \label{de_density}
   \frac{\rho_{\rm de}}{\rho_{\rm de0}} = \exp \left( -3 \int_{1}^a \frac{1+w_{\rm de}(a')}{a'}\, da' \right),
    \end{align}
where $\rho_{\rm de0}$ is the present DE density. It is clear from eq.~\ref{de_density} that DE density evolution varies depending on the functional form of $w_{\rm de}$. In this work, we consider a new parametrization of DE EoS parameter proposed in~\cite{singh2024new} which is given below 
\begin{align}\label{EoS}
w_{\rm de}(a) = -1 + \frac{a^{-\alpha} e^{-2a\alpha} \alpha \, \left[\arctan(a)^{-\alpha}\right]}{3\left(1 + a^{-2\alpha}\right)},
\end{align}
where $\alpha$ is a free parameter to be fixed from cosmological observations. The eq.~\ref{EoS} can be rewritten in terms of redshift $z$ as follows
\begin{align}\label{EoS interms of Z}
w_{\rm de}(z) = -1 + \frac{(1 + z)^\alpha \exp \left(\frac{-2 \,\alpha }{1+z}\right) \, \alpha \, \left[\arctan(1+z)^{\alpha}\right]}{3\left(1 + (1 + z)^{2\alpha}\right)}.
\end{align}
If the parameter $\alpha=0$, this parametrization reduces to $w_{\rm de}(z) = -1$, representing the cosmological constant and hence recovers the $\Lambda$CDM paradigm. However, in the general case, $\alpha \neq 0$ indicates the deviation from the standard $\Lambda$CDM scenario.

\subsection{Physical Interpretation of the Parameter \texorpdfstring{$\alpha$}{alpha}}
The parameter $\alpha$ plays a central role in governing the redshift evolution of the dark energy equation of state in the proposed parametrization. Although introduced as a phenomenological parameter, its sign and magnitude directly control the deviation of the model from the standard LCDM scenario and therefore possess clear theoretical implications. 
The sign of  $\alpha$ determines whether dark energy behaves as a standard scalar field (quintessence) or exhibits phantom evolution. For $\alpha > 0$, the model exhibits a quintessence-like behavior, where $w(z) > -1$. Conversely, $\alpha < 0$ corresponds to a phantom-like regime, where $w(z)$ may cross below $-1$.
The magnitude of $\alpha$ regulates how strongly the dark energy equation of state evolves with redshift. Small values of $|\alpha|$ yield weak deviations from $w = -1$. In contrast, larger magnitudes enhance the contributions of both the power-law and exponential factors in the parametrization, allowing for a more pronounced redshift dependence.

Using eq.~\ref{de_density} and eq.~\ref{EoS interms of Z}, the evolution of the DE density is given by
\begin{align}\label{DE Evolution}
\frac{\rho_{\rm de}}{\rho_{\rm de0}} = \exp \left(\frac{\alpha z}{1+z}\right) \, \left[ \frac{\arctan(1+z)^{\alpha}}{\arctan 1}\right].
\end{align} 
By putting the above parametrization into the first Friedmann equation, we obtained the Hubble parameter as a function of redshift z as follows:
\begin{equation}\label{Novel Model}
  E(z) =\frac{H(z)}{H_0}= \sqrt{\left[\Omega_{\rm m0}\,(1 + z)^3 + \Omega_{\rm d e0}\, \exp \left(\frac{\alpha z}{1+z}\right) \, \frac{\left[\arctan(1+z)^{\alpha}\right]}{\arctan 1}\right]},
\end{equation}
where $H_0$ is the present value of the Hubble parameter; $\mathrm{\Omega}_{\rm m0} =\frac{8 \pi G}{3H^2}\rho_{\rm m0}$ and $\mathrm{\Omega}_{\rm de0} = \frac{8 \pi G}{3H^2}\rho_{\rm de0}$ are respectively, the matter and DE density parameter at present. This is the actual expression, which enables us to examine the cosmological evolution in depth and test it using observational probes.

\section{Data Sets and Methodology} \label{model_data}
In this section, we explain the details of the data sets used and the statistical technique for constraining the parameters of the derived model. The primary objective is to assess the model's consistency by comparing theoretical predictions with observations.

\subsection{\textbf{\textit{Observational Hubble Data (OHD)}}}
We can construct the Hubble parameter directly using the Cosmic Chronometer (CC) method by studying the oldest and passively evolving galaxies by using different techniques. Here, we employ 66 OHD points obtained form CC technique with a wide range of $0.07\leq z\leq 1.965$ which are represented in Table~\ref{OHD data} with corresponding sources. 
The $\chi^2$-function for OHD can be expressed as follows: 
\begin{align}\label{chi^2}
\chi^2 = \sum_{j=1}^{66} \left[\frac{H_{\text{th}}\left(\alpha, H_0, z_j\right) - H_{\text{obs}}\left(z_j\right)}{\sigma_H\left(z_j\right)}\right]^2,
\end{align}
where $H_{\text{th}}\left(\alpha, H_0, z_j\right)$ is the theoretical value obtained from the model based on various $z_j$ and a parameter $\alpha$, $H_{\text{obs}}\left(z_j\right)$ is the observed value and $\sigma_H$ represents the error term in the observed Hubble parameter.

\begin{table}[ht]
\tbl{The 66 OHD measurements in $\mathrm{Km\,s^ {-1} Mpc^{-1}}$ used in this study, obtained via cosmic chronometers.\label{OHD data}}
{
\scriptsize
\begin{minipage}{0.48\linewidth}
\begin{tabular}{@{}cccc@{}} \toprule
S.No & $z$  & $H(z)$            & Ref. \\ \colrule
1 & 0.00    & 69.1$ \pm $ 1.3   &~\cite{Farooq:2016zwm}\\  
2 & 0.07    & 70.4$ \pm $ 20    &~\cite{zhang2016test} \\  
3 & 0.07    & 69.0$ \pm $ 19.6  &~\cite{zhang2016test} \\  
4 & 0.09    & 70.4$ \pm $ 12.2  &~\cite{simon2005constraints}\\  
5 & 0.10    & 70.4$ \pm $ 12.2  &~\cite{zhang2016test}\\  
6 & 0.120   & 68.6$ \pm $ 26.2  &~\cite{zhang2016test}\\   
7 & 0.12    & 70.0$ \pm $ 26.7  &~\cite{Farooq:2016zwm}\\   
8 & 0.170   & 83.0$ \pm $ 8     &~\cite{simon2005constraints}\\  
9 & 0.170   & 84.7$ \pm $ 8.2   &~\cite{simon2005constraints}\\  
10 & 0.179  & 76.5$ \pm $ 4     &~\cite{Moresco:2012jh}\\  
11 & 0.1791 & 75.0$ \pm $ 4     &~\cite{Moresco:2012jh}\\  
12 & 0.199  & 76.5$ \pm $ 5.1   &~\cite{Moresco:2012jh}\\  
13 & 0.1993 & 75.0$ \pm $ 5     &~\cite{Moresco:2012jh}\\  
14 & 0.200  & 72.9$ \pm $ 29.6  &~\cite{zhang2016test}\\  
15 & 0.20   & 74.4$ \pm $ 30.2  &~\cite{zhang2016test}\\  
16 & 0.27   & 78.6$ \pm $ 14.3  &~\cite{simon2005constraints}\\  
17 & 0.280  & 88.8$ \pm $ 36.3  &~\cite{zhang2016test}\\  
18 & 0.28   & 90.6$ \pm $ 37.3  &~\cite{Farooq:2016zwm}\\   
19 & 0.3519 & 83.0$ \pm $ 14    &~\cite{Moresco:2012jh}\\    
20 & 0.352  & 84.7$ \pm $ 14.3  &~\cite{Moresco:2012jh}\\   
21 & 0.3802 & 83.0$ \pm $ 13.5  &~\cite{moresco20166}\\ 
22 & 0.3802 & 84.7$ \pm $ 14.1  &~\cite{moresco20166}\\
23 & 0.40   & 95.0$ \pm $ 17    &~\cite{simon2005constraints}\\    
24 & 0.40   & 96.9$ \pm $ 17.3  &~\cite{simon2005constraints}\\  
25 & 0.4004 & 77.0$ \pm $ 10.2  &~\cite{moresco20166}\\
26 & 0.4004 & 78.6$ \pm $ 10.4  &~\cite{moresco20166}\\
27 & 0.4247 & 87.1$ \pm $ 11.2  &~\cite{moresco20166}\\ 
28 & 0.4247 & 88.9$ \pm $ 11.4  &~\cite{moresco20166}\\ 
29 & 0.43   & 88.3$ \pm $ 3.8   &~\cite{Farooq:2016zwm}\\   
30 & 0.44   & 84.3$ \pm $ 7.9   &~\cite{blake2012wigglez}\\ 
31 & 0.4497 & 92.8$ \pm $ 12.9  &~\cite{moresco20166}\\ 
32 & 0.4497 & 94.7$ \pm $ 13.1  &~\cite{moresco20166}\\  
33 &  0.470 & 89.0$ \pm $ 34.0  &~\cite{Ratsimbazafy:2017vga}\\ 
\botrule
\end{tabular}
\end{minipage}
\hfill
\begin{minipage}{0.48\linewidth}
\begin{tabular}{@{}cccc@{}} \toprule
S.No & $z$ & $H(z)$ & Ref. \\ \colrule
34 & 0.47   & 90.8$ \pm $ 50.6  &~\cite{Ratsimbazafy:2017vga}\\ 
35 & 0.4783 & 80.0$ \pm $ 99.0  &~\cite{moresco20166}\\
36 & 0.4783 & 82.5$ \pm $ 9.2   &~\cite{moresco20166}\\  	
37 & 0.48   & 99.0$ \pm $ 63.2  &~\cite{Ratsimbazafy:2017vga}\\   
38 & 0.593  & 104.0$ \pm $ 13.0 &~\cite{Moresco:2012jh}\\    
39 & 0.593  & 106.1$ \pm $ 13.3 &~\cite{Moresco:2012jh}\\   
40 & 0.60   & 89.7$ \pm $ 6.2   &~\cite{zhang2016test}\\  
41 & 0.6797 & 92.0$ \pm $ 8     &~\cite{Moresco:2012jh}\\    
42 & 0.68   & 93.9$ \pm $ 8.1   &~\cite{Moresco:2012jh}\\  
43 & 0.73   & 99.3$ \pm $ 7.1   &~\cite{blake2012wigglez}\\ 
44 & 0.7812 & 105.0$ \pm $ 12   &~\cite{Moresco:2012jh}\\       
45 & 0.781  & 107.1$ \pm $ 12.2 &~\cite{Moresco:2012jh}\\    
46 & 0.875  & 127.6$ \pm $ 17.3 &~\cite{Moresco:2012jh}\\    
47 & 0.8754 & 125.0$ \pm $ 17   &~\cite{Moresco:2012jh}\\    
48 & 0.88   & 91.8$ \pm $40.8   &~\cite{stern2010cosmic}\\  
49 & 0.880  & 90.0$ \pm $ 40    &~\cite{Ratsimbazafy:2017vga}\\ 
50 & 0.90   & 69.0 $ \pm $ 12   &~\cite{simon2005constraints}\\   
51 & 0.90   & 119.4$ \pm $ 23.4 &~\cite{simon2005constraints}\\
52 & 0.900  & 117.0$ \pm $ 23   &~\cite{simon2005constraints}\\ 
53 & 1.037  & 157.2$ \pm $ 20.4 &~\cite{Moresco:2012jh}\\    
54 & 1.037  & 154.0$ \pm $ 20   &~\cite{Moresco:2012jh}\\   
55 & 1.30   & 171.4$ \pm $ 17.3 &~\cite{simon2005constraints}\\ 
56 & 1.300  & 168.0$ \pm $ 17   &~\cite{simon2005constraints}\\
57 & 1.363  & 160.0$ \pm $ 33.6 &~\cite{moresco2015raising}\\  
58 & 1.363  & 163.3$ \pm $ 34.3 &~\cite{moresco2015raising}\\ 
59 & 1.430  & 177.0$ \pm $ 18   &~\cite{simon2005constraints}\\   
60 & 1.43   & 180.6$ \pm $ 18.3 &~\cite{simon2005constraints}\\  
61 & 1.530  & 140.0$ \pm $ 14   &~\cite{simon2005constraints}\\  
62 & 1.53   & 142.9$ \pm $ 14.2 &~\cite{simon2005constraints}\\ 
63 & 1.750  & 202.0$ \pm $ 40   &~\cite{simon2005constraints}\\  
64 & 1.75   & 206.1$ \pm $ 40.8 &~\cite{simon2005constraints}\\ 
65 & 1.965  & 186.5$ \pm $ 50.4 &~\cite{moresco2015raising}\\
66 & 1.965  & 190.3$ \pm $ 51.4 &~\cite{moresco2015raising}\\
\botrule
\end{tabular}
\end{minipage}
}
\end{table}

\subsection{\textbf{\textit{Pantheon+ and SH0ES}}}
Type Ia supernovae (SNe Ia) are commonly used as standard candles due to their uniform peak luminosity. We employ the distance modulus measurements collected from the SNe Ia. The apparent brightness $m_B$ of a supernova observed at redshift $z$ reads:
\begin{eqnarray}
\label{distance_modulus}
m_B - M_B = 5 \log_{10} \left[ \frac{D_L(z)}{1\rm Mpc} \right] + 25\, ,
\end{eqnarray}
where $M_B$ denotes the absolute magnitude, and $D_L(z)$ denotes the luminosity distance given by
\begin{equation}
    D_L(z)=\frac{\left(1+z\right)}{H_0}\int_{0}^{z}\frac{1}{E(z')} dz',
\end{equation}
where $E(z')$ can be obtained from eq.~\ref{Novel Model} by replacing $z'$ as $z$.

We chose 1701 light curves that correspond to 1550 different SNe Ia in the redshift range of $z \in [0.01, 2.26]$ based on SNe Ia distance modulus data from the Pantheon+ sample~\cite{brout2022pantheon+}. The SH0ES Cepheid host distance anchors~\cite{brout2022pantheon+} are also included in our analysis. This combined dataset is hereafter referred to as PPS. 
These measurements have been considered from 18 different surveys including CfA1-CfA4\cite{Riess:1998dv, Jha:2005jg, Hicken:2009df, Hicken:2012zr}, CSP~\cite{krisciunas2017carnegie}, SOUSA~\cite{brown2014sousa}, CNIa0.02~\cite{Chen:2020qnp}, Foundation~\cite{foley2018foundation}, and LOSS~\cite{stahl2019lick} which are focus on the redshift range of $0.01$ to $0.1$. Further,  DES~\cite{brout2019first}, SNLS~\cite{SDSS:2014iwm}, SDSS~\cite{Sako:2011um}, PS1~\cite{Pan-STARRS1:2017jku} have focused on $z \geq 0.1$ and SCP, GOODS, HDFN, CANDLES/CLASH have released the data for $z$ greater than $1.0$~\cite{SupernovaSearchTeam:2004lze, Riess:2006fw, SupernovaCosmologyProject:2011ycw}. The $\chi^2$-function for PPS can be expressed as follows:
\begin{eqnarray}
\label{chi^2 for PPS}
\chi^2=\sum_{j,k=1}^{1701}{\mathrm{\nabla}_{\mu_j}\left(C_{SN}^{-1}\right)_{j,k}}\mathrm{\nabla}_{\mu_k},
\end{eqnarray}
where $\mathrm{\nabla}_\mu=\mu_{\rm th}-\mu_{\rm obs}$ quantifies the deviation between theoretical and observed distance modulus values. The covariance matrix $C_{SN}$ consists of systematic and statistical errors which can be obtained from~\cite{PPSdata}. Theoretically, the distance modulus function is calculated using
\begin{equation}
    \mu_{\rm th}=5 \log{\left[D_L(z)\right]}+25\, .
\end{equation}

\subsection{\textbf{\textit{DESI BAO DR2}}}
In the density of visible baryonic matter, recurring and periodic fluctuations occur which are commonly known as BAO. These oscillations serve as crucial standard rulers for accurately measuring distances in the field of cosmology. The BAO scale can be expressed as the sound horizon at the drag epoch which is the distance a sound wave would have traveled before baryon-photon decoupling. The drag epoch occurs after photon decoupling because of the residual interaction of baryons with photons, as baryons were significantly fewer in number compared to photons. The drag scale $r_d$ is determined as follows:
\begin{equation}
  r_d = \int_{z_d}^{\infty} \frac{c_s(z)}{H(z)} \, dz,
\end{equation}
where $c_s(z)$ denotes baryon-photon plasma’s sound speed and $z_d$ represents the redshift at the drag epoch.

The DESI is an advanced spectroscopic survey to improve BAO constraints by tracing the large-scale matter distribution.  
The DESI measures redshifts of approximately 30 million galaxies and quasars, significantly enhancing our constraints on the cosmological model. This survey has two main observing programs, depending on the night-time sky conditions: the bright and dark programs. The Bright Galaxy Sample (BGS) is observed under bright-sky conditions, typically around the full moon, whereas the dark-sky observations focus on Emission Line Galaxies (ELGs), Luminous Red Galaxies (LRGs), and Quasi-Stellar Objects (QSOs). In particular, Lyman-alpha (Ly-$\alpha$) forest measurements, which probe the intergalactic medium at high redshifts (typically $z>2$), are obtained from QSO spectra. These are characterized in terms of measurements of the Hubble horizon ($D_H/r_d$), transverse comoving distance ($D_M/r_d$), and the angle-averaged distance ($D_V/r_d$), normalized to the comoving acoustic horizon at the drag epoch. In the second data release (DR2), DESI has achieved substantial observational coverage, comprising 6671 dark tiles and 5171 bright tiles. This represents an increase of approximately 2.4 times for the dark program and 2.3 times for the bright program compared to the first data release (DR1)~\cite{adame2025desi}. Using DESI BAO DR2 measurements, one can determine the product ${H_0}{r_d}$, which serves as an essential cosmological parameter but does not constrain $H_0$  or $r_d$ independently. In this work, we have considered DESI BAO DR2 observations from~\cite{karim2025desi} as mentioned in Table~\ref{DESI BAO}.

\begin{table}[ht]
\tbl{The 9 BAO measurements from DESI DR2 used in this analysis, expressed in units of $\mathrm{Km\,s^{-1}\,Mpc^{-1}}$.\label{DESI BAO}}
{\begin{tabular}{@{}ccccc@{}} \toprule
{tracer} &  $z_{\rm eff}$ & $D_M/r_d$ & $D_H/r_d$ & $D_V/r_d$ \\
\colrule
                {\scriptsize {BGS }}  &  $0.295$ & $-$ &$-$ & $7.942 \pm 0.075$
				\\
				{\scriptsize {LRG1}}  &  $0.510$ & $13.588 \pm 0.167$ & $ 21.863 \pm 0.425$ & $12.720 \pm 0.099$
				\\
				{\scriptsize {LRG2}} &  $0.706$ & $17.351 \pm 0.177$ & $19.455 \pm 0.330$ & $16.050 \pm 0.110$
				\\
				{\scriptsize {LRG3} } &  $0.922$ & $21.648 \pm 0.178$ & $17.577 \pm 0.213$ & $19.656 \pm 0.105$
                \\
                {\scriptsize {ELG1} } &  $0.955$ & $21.707 \pm 0.335$ & $17.803 \pm 0.297$ & $20.008 \pm 0.183$
                \\
                {\scriptsize {LRG3+ELG1} } &  $0.934$ & $21.576 \pm 0.152$ & $17.641 \pm 0.193$ &$19.721 \pm 0.091$
                \\
                
                {\scriptsize {ELG2} } &  $1.321$ & $27.601 \pm 0.318$ & $14.176 \pm 0.221$ & $24.252 \pm 0.174$
                \\
                {\scriptsize {QSO} } & $1.484$ & $30.512 \pm 0.760$ &$12.817 \pm 0.516$ & $26.055 \pm 0.398$ 
                \\
                {\scriptsize {Ly-$\alpha$ QSO} } &  $2.330$ & $38.988 \pm 0.531$ & $8.632 \pm 0.101$ & $31.267 \pm 0.256$
				\\
                \botrule
\end{tabular}}
\end{table}

We have performed MCMC analysis using the \texttt{emcee} Python package~\cite{foreman2013emcee} to get the sample on the model parameters.
To examine MCMC samples, we used the \texttt{GetDist} Python package~\cite{lewis2019getdist}. 
We have used flat priors in a broader range as compared to~\cite{singh2024new} for a better exploration of model behavior. The priors are mentioned as: $H_0 \in [40,100]$, $\Omega_{m0} \in [0.01,0.6]$, and $\alpha \in [-1,2]$ on the model parameter.

\section{Results and Discussion}\label{results}
\begin{figure}[ht]
    \centerline{\includegraphics[scale=0.5]{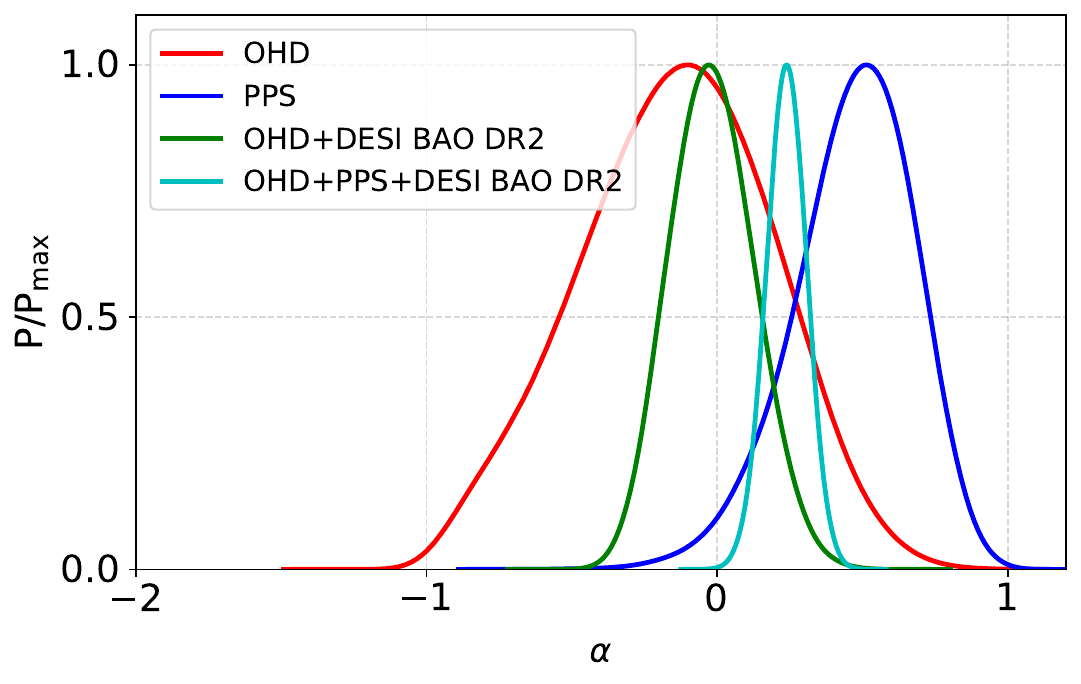}}
	\caption{ One-dimensional marginalized probability distribution for $\alpha$.}
		\label{alpha vs Pmax/p_max}
\end{figure}

We have presented the derived constraints on the parameters of the derived model in Table~\ref{params} from four datasets: OHD, PPS, OHD + DESI BAO DR2, and OHD + PPS + DESI BAO DR2. We have noted that the parameter $\alpha$ is consistent with zero with OHD and OHD + DESI BAO DR2 whereas it has non-null values for PPS and joint case: OHD + PPS + DESI BAO DR2. The higher mean value of $\alpha$ is obtained from the PPS dataset with $\alpha=0.47^{+0.24}_{-0.17}$ at 68\% CL. The tightest bound of parameter $\alpha$ is obtained from joint analysis with $\alpha =0.239 \pm 0.07$ at 68\% CL. These constraints on parameter $\alpha$ vary across the dataset and show the constraining power of the OHD, PPS, and their combinations with DESI BAO DR2 rather than almost the same mean value as obtained in~\cite{singh2024new}. The Fig.~\ref{alpha vs Pmax/p_max} shows one-dimensional marginalized probability distributions of parameter $\alpha$ for all cases. We emphasize that the deviation of $\alpha$ from zero indicates the dynamical nature of DE. In the joint analysis, a clear deviation of $\alpha$ from zero is observed, indicating the preference for the dynamical nature of DE. The preferred positive values of $\alpha$ across PPS and OHD+PPS+DESI BAO DR2 datasets indicate a quintessence-like behavior of DE.

\begin{table}[ht]
\tbl{Summary of the observational constraints on the parameter values from datasets.\label{params}}{
\begin{tabular}{@{}lcccc@{}}
\toprule
Dataset & Model & $H_0$  & $\Omega_{m0}$ & $\alpha$ \\
\midrule

\multirow{2}{*}{OHD} 
 & Our Model & $63.3\pm 1.3$ & $0.365\pm 0.052$ & $-0.15^{+0.37}_{-0.32}$ \\
 & {\color{blue}{$\Lambda$CDM}} & {\color{blue}{$69.2 \pm 1.1$}} & {\color{blue}{$0.295^{+0.024}_{-0.027}$}} & {\color{blue}{$-$}} \\
\midrule

\multirow{2}{*}{PPS} 
 & Our Model & $69.26\pm 0.83$ & $0.282\pm 0.072$ & $0.47^{+0.24}_{-0.17}$ \\
 & {\color{blue}{$\Lambda$CDM}} & {\color{blue}{$72.84 \pm 0.23$}} & {\color{blue}{$0.362 \pm 0.019$}} & {\color{blue}{$-$}} \\
\midrule

\multirow{2}{*}{OHD+DESI BAO DR2} 
 & Our Model & $67.20\pm 0.82$ & $0.3395\pm 0.0088$ & $-0.02^{+0.14}_{-0.16}$ \\
 & {\color{blue}{$\Lambda$CDM}} & {\color{blue}{$69.05 \pm 0.83$}} & {\color{blue}{$0.2985 \pm 0.0078$}} & {\color{blue}{$-$}} \\
\midrule

\multirow{2}{*}{OHD+PPS+DESI BAO DR2} 
 & Our Model & $68.40\pm 0.23$ & $0.3337\pm 0.0086$ & $0.239\pm 0.070$\\
 & {\color{blue}{$\Lambda$CDM}} & {\color{blue}{$73.38 \pm 0.15$}} & {\color{blue}{$0.3007 \pm 0.007$}} & {\color{blue}{$-$}} \\
\bottomrule
\end{tabular}}
\end{table}

Our model accommodates varying mean values of the Hubble constant in the range $H_0 \approx 63-69\, \mathrm{Km\,s^ {-1} Mpc^{-1}}$ depending on the dataset combination used. Our model provides the Hubble constant, $H_0=63.30 \pm 1.3\, \mathrm{Km\,s^ {-1} Mpc^{-1}}$, $H_0=69.26 \pm 0.83\, \mathrm{Km\,s^ {-1} Mpc^{-1}}$, $H_0=67.20 \pm 0.82\, \mathrm{Km\,s^ {-1} Mpc^{-1}}$, and $H_0=68.40 \pm 0.23\, \mathrm{Km\,s^ {-1} Mpc^{-1}}$ at 68\% CL from the OHD, PPS, OHD + DESI BAO DR2, and OHD + PPS + DESI BAO DR2, respectively. To examine the efficacy of our model results, we have derived constraints on the standard $\Lambda$CDM model with the same dataset combinations. The $\Lambda$CDM model predicts Hubble constant $H_0$ in the range from $H_0 \approx 69\, \mathrm{Km\,s^ {-1} Mpc^{-1}}$ to $H_0 \approx 73\, \mathrm{Km\,s^ {-1} Mpc^{-1}}$.
We have noticed that our model systematically prefers lower $H_0$ values than the $\Lambda$CDM model with all data combinations, Table~\ref{params}. From the OHD alone, our model yields $H_0=63.3 \pm 1.3$, which is significantly lower than the $\Lambda$CDM model estimate $H_0=69.2 \pm 1.1$ at 68\% CL. This low $H_0$ arises from the additional freedom in the parametrization rather than from the OHD compilation itself.
From OHD + DESI BAO DR2, the estimate $H_0=67.20 \pm 0.82\, \mathrm{Km\,s^ {-1} Mpc^{-1}}$ is more aligned with the early universe measurements such as from the Planck collaboration which yeilds $H_0=67.37 \pm 0.54\, \mathrm{Km\,s^ {-1} Mpc^{-1}}$ ~\cite{Planck:2018vyg} and
recent ground based SPT-3G measurement which yield $H_0 = 66.66 \pm 0.60\, \mathrm{Km\,s^ {-1} Mpc^{-1}}$~\cite{SPT-3G:2025bzu}.

From PPS alone and joint case, our model provides $H_0=69.26 \pm 0.83$ (PPS) and $H_0=68.40 \pm 0.23$ (joint case), both at 68\% CL, (Table~\ref{params}). The $\Lambda$CDM model yields $H_0=72.84 \pm 0.23$ (PPS) and $H_0=73.38 \pm 0.15$ (joint case), both are in excellent agreement with local SH0ES measurement, $H_0 = 73.04 \pm 1.04$ $\mathrm{Km\,s^{-1} Mpc^{-1}}$~\cite{SupernovaSearchTeam:2004lze}.
Our combined data constraints on the Hubble constant lie above the early universe-CMB measurement and below the late time SHOES measurement. This shift in $H_0$ towards the intermediate value is due to the re-distribution of the expansion history relative to the $\Lambda$CDM, due to the extra parameter $\alpha$. Note that we have reported for the PPS and joint case, the parameter $\alpha$ deviates from zero, implying a statistically significant departure from the $\Lambda$CDM. Based on our results on $H_0$, we can not significantly claim to resolve the Hubble tension, but our findings contribute to the debate on the Hubble constant tension, as discussed in Section \ref{sec:intro}.

\begin{figure}[ht]
 \centerline{ \includegraphics[scale=0.85]{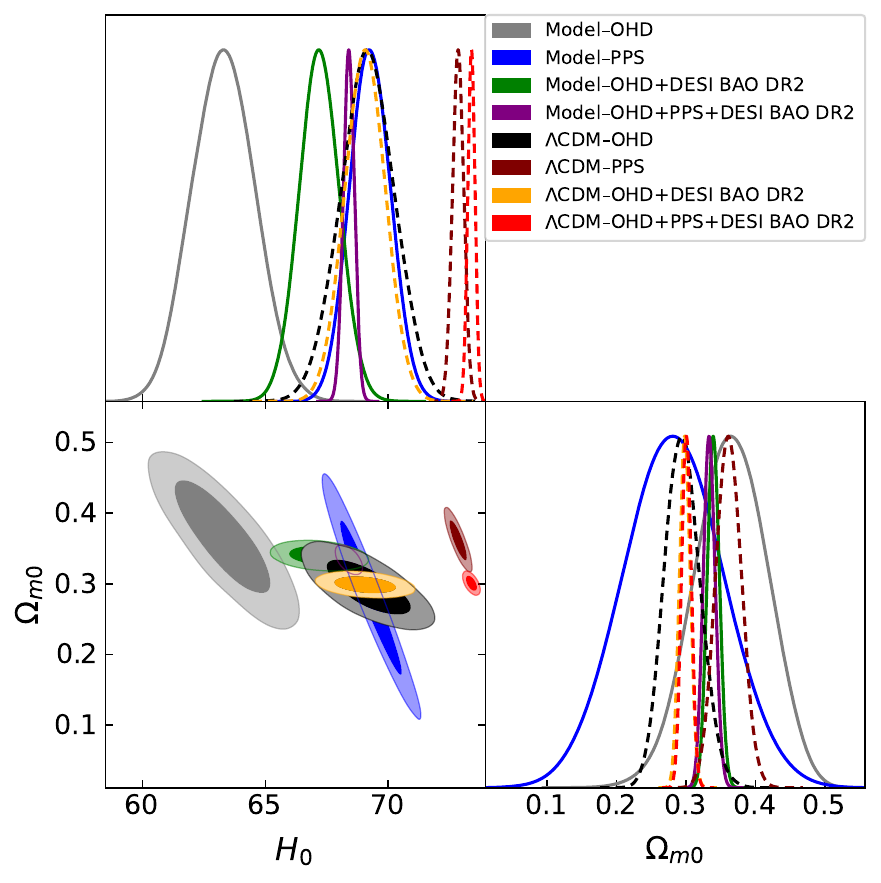}}
	\caption{{One and two-dimensional confidence regions at 68\% and 95\% CL for the parameters $H_0$ and $\Omega_{m0}$ for our model and $\Lambda$CDM  model.}} 
\label{combined}
\end{figure}

The posterior distribution of some parameters of our model and $\Lambda$CDM model at 68\% CL and 95\% CL are displayed in Fig.~\ref{combined} from all data combinations. Overall, the results obtained in the present work are more robust and precise than those reported in~\cite{singh2024new} due to the consideration of broader prior ranges, in light of the recent and advance datasets used. 
The findings on matter density parameter, $\Omega_{m0}$ in all cases are consistent with the Planck CMB measurement based on the $\Lambda$CDM model, see the Table~\ref{params}.

\begin{figure}[ht]
 \centerline{ \includegraphics[scale=0.5]{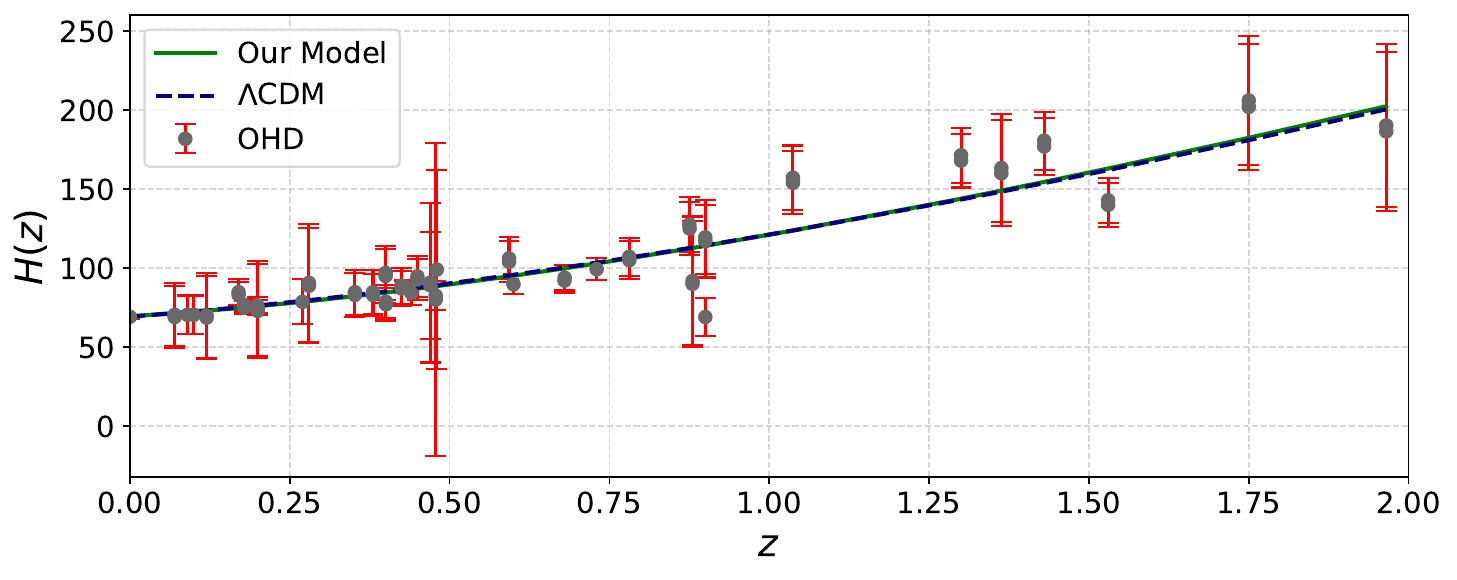}}
	\caption{The error bar plot of OHD points. The solid green and black dotted curves represent the Hubble function for our model and the $\Lambda$CDM model, respectively.}
\label{Model fitting with data points1}
\end{figure}

\begin{figure}[ht]
 \centerline{ \includegraphics[scale=0.5]{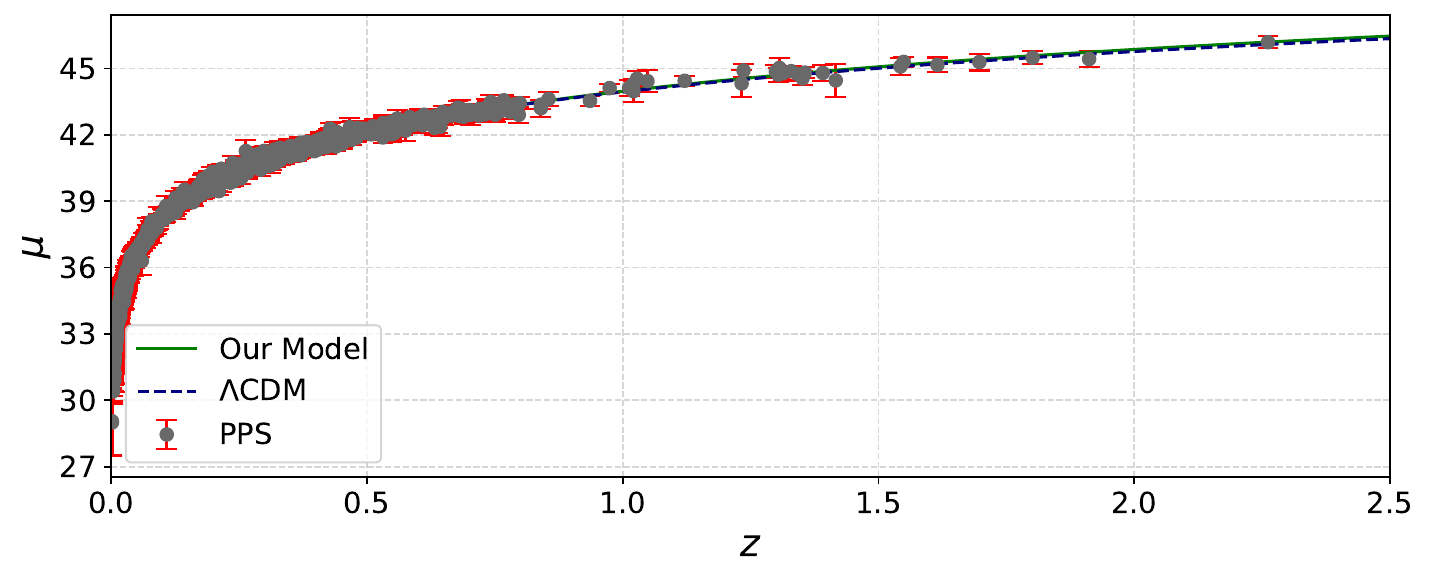}}
	\caption{The error bar plot of 1701 PPS observations. The solid green and black dotted curves represent the distance modulus function for our model and the $\Lambda$CDM model, respectively. }
\label{Model fitting with data points2}
\end{figure}

The comparative fit of the considered model and the standard $\Lambda$CDM model from OHD is shown in Fig.~\ref{Model fitting with data points1} for the redshift from $z=0$ to $z=2.5$ where the OHD points are displayed with their corresponding error bars in red color. It can be seen that both models are consistent with the data at low redshifts whereas the considered model shows a better fit with OHD at high redshifts as compared to the $\Lambda$CDM framework. 
The Fig.~\ref{Model fitting with data points2} represents the distance modulus $\mu(z)$ versus redshift plot from PPS observations where it can be seen that both models fit well with the data points even at high redshifts.

\begin{table}[ht]
\tbl{Summary of $\chi^2_{min}$, $\chi^2_r$, AIC, and BIC values from datasets.\label{chi}}{
\begin{tabular}{@{}lccccc@{}} 
\toprule
Dataset  & Model & $\chi^2_{min}$ & $\chi^2_r$ & AIC & BIC \\ \colrule
\multirow{2}{*}{OHD} & Our Model  & $45.09$ & $0.62$ & $51.09$ & $57.66$ \\
& \textcolor{blue}{$\Lambda$CDM}  & \textcolor{blue}{$45.20$} & \textcolor{blue}{$0.61$} & \textcolor{blue}{$49.20$} & \textcolor{blue}{$53.58$}\\
\hline
\multirow{2}{*}{PPS} & Our Model & $1748.14$ & $1.03$ & $1754.14$ & $1770.46$ \\
& \textcolor{blue}{$\Lambda$CDM} & \textcolor{blue}{1752.48} & \textcolor{blue}{$1.03$} & \textcolor{blue}{$1756.48$} & \textcolor{blue}{$1767.35$} \\
\hline
\multirow{2}{*}{OHD+DESI BAO DR2} & Our Model  & $80.90$ & $1.14$ & $88.90$ & $98.17$ \\
& \textcolor{blue}{$\Lambda$CDM} & \textcolor{blue}{$61.70$} & \textcolor{blue}{$0.86$} & \textcolor{blue}{$67.70$} & \textcolor{blue}{$74.65$}\\
\hline
\multirow{2}{*}{OHD+PPS+DESI BAO DR2} & Our Model  & $1836.85$ & $1.04$ & $1844.85$ & $1866.78$ \\
& \textcolor{blue}{$\Lambda$CDM} & \textcolor{blue}{$1855.80$} & \textcolor{blue}{$1.05$} & \textcolor{blue}{$1861.80$} & \textcolor{blue}{$1878.25$}\\

\botrule
\end{tabular}}
\end{table}

\subsection{Comparison with other studies}
In this section, we compare the findings of the present work with other recent investigations as discussed in the literature using different datasets. The authors in~\cite{Tripathi:2016slv} investigated both constant and time-varying EoS parametrizations with SNe Ia, BAO, and H(z) measurements. They have reported a matter density $\Omega_{m0} \approx 0.28$ for the $w$CDM model and similar values in the range $0.28-0.30$ for dynamical DE models. They have fixed the Hubble parameter to $H_0=70\,\mathrm{Km\,s^ {-1} Mpc^{-1}}$ which does not allow a direct constraint on its value. Our results allow $H_0$ to vary freely during parameter estimation, which naturally broadens the allowed parameter space. Consequently, we obtain slightly higher values of the matter density parameter, lying in the range $\Omega_{m0} = 0.282-0.365$ depending on the dataset combination used (see Table~\ref{params}). This increase arises primarily because our analysis simultaneously constrains both $\Omega_{m0}$ and $H_0$, rather than fixing $H_0$ as 70 $\mathrm{Km\,s^ {-1} Mpc^{-1}}$.
The authors in~\cite{Chaudhary:2025bfs} investigated some well-known DE parametrizations, including CPL, Logarithmic, Exponential, JBP, BA, and GEDE, using recent data from CC, DESI DR2 BAO, Pantheon+, DES-SN5Y, and Union3. The analysis also presents similar constraints on $H_0$ lying in the range $H_0 \approx66 - 69 \, \mathrm{Km\,s^ {-1} Mpc^{-1}}$ in the joint case (CC+DESI DR2+Pantheon+). Also, they reported $\Omega_{m0} \approx 0.30$ across all models, which aligns closely with the results of the present work. 
Furthermore, the authors in~\cite{Capozziello:2025lor} extended the investigation of cosmological models in~\cite{Chaudhary:2025bfs} by incorporating some additional probes, including the compressed CMB shift parameter. They find that the Hubble constant lies in the range $H_0 \approx 62 - 69 \, \mathrm{Km\,s^ {-1} Mpc^{-1}}$ and present matter density, $\Omega_{m0} \approx 0.30-0.34 $ across different data combinations for considered parametrizations of DE. These findings are closely aligned with our results for the DE parametrization in this work, in light of the dataset used. Further, when comparing from other scientific literature, the Hubble constant values are aligned very closely with recent works such as $H_0=68.05 \pm 0.38\, \mathrm{Km\,s^ {-1} Mpc^{-1}}$ by de Cruz Perez et al.~\cite{deCruzPerez:2024shj}, $H_0=67.96 \pm 0.38\, \mathrm{Km\,s^ {-1} Mpc^{-1}}$ by Adame et al.~\cite{adame2025desi}, $H_0=69.80 \pm 1.7\, \mathrm{Km\,s^ {-1} Mpc^{-1}}$ by Cao et al.~\cite{Cao:2023eja}, and  $H_0=69.56 \pm 0.63\, \mathrm{Km\,s^ {-1} Mpc^{-1}}$ by Roy et al.~\cite{Roy:2024kni}.

\subsection{Model Comparison}
To examine the goodness of fit of our model to the observational datasets, we performed a chi-square ($\chi^2$) analysis to test the consistency of the model with the data used. The minimum chi-square, $\chi^2_{\text{min}}$, can be estimated using the relation $\chi^2_{\text{min}} = -2\ln(\mathcal{L}_{\text{max}})$, where $\mathcal{L}_{\text{max}}$  is the maximum likelihood function. Further, the reduced chi-square is obtained after dividing $\chi^2_{\text{min}}$ by the number of degrees of freedom ($N-k$), where $N$ is the number of data points and $k$ is the number of free model parameters. A value of $\chi^2_r\ \approx\ 1$ typically indicates a satisfactory fit, while significant deviations from this value suggest either under-fitting ($\chi^2_r\ \gg\ 1$) or overfitting ($\chi^2_r\ \ll\ 1$). The computed values  $\chi^2_{\text{min}}$ and $\chi^2_r$ are listed in Table~\ref{chi}. All combinations of data provide $\chi^2_r$ values very close to 1, showing overall consistency, except in the case of OHD  and OHD + DESI BAO DR2 where a marginal sign of statistical underfitting and overfitting is observed. 

We also performed the statistical performance using Akaike Information Criterion (AIC)~\cite{akaike2003new,burnham2011aic} and Bayesian Information Criterion (BIC)~\cite{schwarz1978estimating} for our model with the $\Lambda$CDM model. These criteria provide an assessment for model selection by penalizing additional free parameters to prevent overfitting. The AIC and BIC are calculated as $\mathrm{AIC} = \chi^2_{\text{min}} + 2k$, and $\mathrm{BIC} = \chi^2_{\text{min}} + k\ln(N)$. The statistical performance between two models can be computed using $\Delta \mathrm{AIC} = \mathrm{AIC}_{\text{model}} - \mathrm{AIC}_{\Lambda\text{CDM}}$, and $\Delta \mathrm{BIC} = \mathrm{BIC}_{\text{model}} - \mathrm{BIC}_{\Lambda\text{CDM}}$. As noted in~\cite{Tan:2011pa}, a model may be regarded as significantly preferred if the AIC difference exceeds a threshold value of 5. In standard model selection, a threshold of $\Delta \mathrm{AIC}$ or $\Delta \mathrm{BIC} \ge 5$ indicates a strong preference for the model with a lower value.
The obtained AIC and BIC values for the considered models are summarized in Table~\ref{chi}. For the OHD data alone, we obtain $\Delta\mathrm{AIC} \approx 1.89$, and $\Delta\mathrm{BIC}\approx 4.08$, suggesting that both models are statistically comparable, with $\Lambda$CDM being mildly favored under BIC. For the PPS dataset, AIC prefers our model with $\Delta\mathrm{AIC} \approx -2.34$, whereas BIC favors $\Lambda$CDM with $\Delta\mathrm{BIC}\approx 3.11$. Combining OHD with the DESI BAO DR2 dataset strongly favors the $\Lambda$CDM model with $\Delta\mathrm{AIC} \approx 21.20$, and $\Delta\mathrm{BIC}\approx 23.52$. We obtain $\Delta\mathrm{AIC} \approx -16.95$, and $\Delta\mathrm{BIC}\approx -11.47$, which strongly favors our model under OHD+PPS+DESI BAO DR2 dataset.

\section{Physical Behavior of the Model}\label{model_properties}

\subsection{\textbf{\textit{Deceleration parameter}}}
The deceleration parameter is an important quantity to understand the decelerated or accelerated behavior of the universe. It is defined as $ q = -1-\frac{\dot{H}} {{H^{2}}}$, where overdot denotes derivative with respect to time. The sign of $q$ serves as an indicator of whether the cosmic expansion is accelerating or decelerating. A positive value ($q > 0$) indicates a decelerating phase of the universe, as it was in the matter and radiation dominated epoch, whereas a negative ($q < 0$) corresponds to an accelerating phase, as the current epoch of the universe is dominated by DE. The deceleration parameter for the derived model reads as:

\begin{equation}\label{deceleratione eq}
  q(z) = -1+\frac{\Big[3(1+z)^4\Omega_{m}+\frac{4\alpha\Omega_{de}(1+z)^2 e^\frac{\alpha z}{1+z}\arctan (1+z)^{\alpha-1}}{\pi(z^2+2z+2)}+\frac{4 \alpha \Omega_{de} e^\frac{\alpha z}{1+z}\arctan (1+z)^{\alpha}}{\pi}\Big]}{2(1+z)\Big[(1+z)^3 \Omega_{m}+\frac{4 \Omega_{de}e^\frac{\alpha z}{1+z}\arctan (1+z)^{\alpha}}{\pi}\Big]},
\end{equation}

\begin{figure} [ht]
\centerline	 {\includegraphics[scale=0.5]{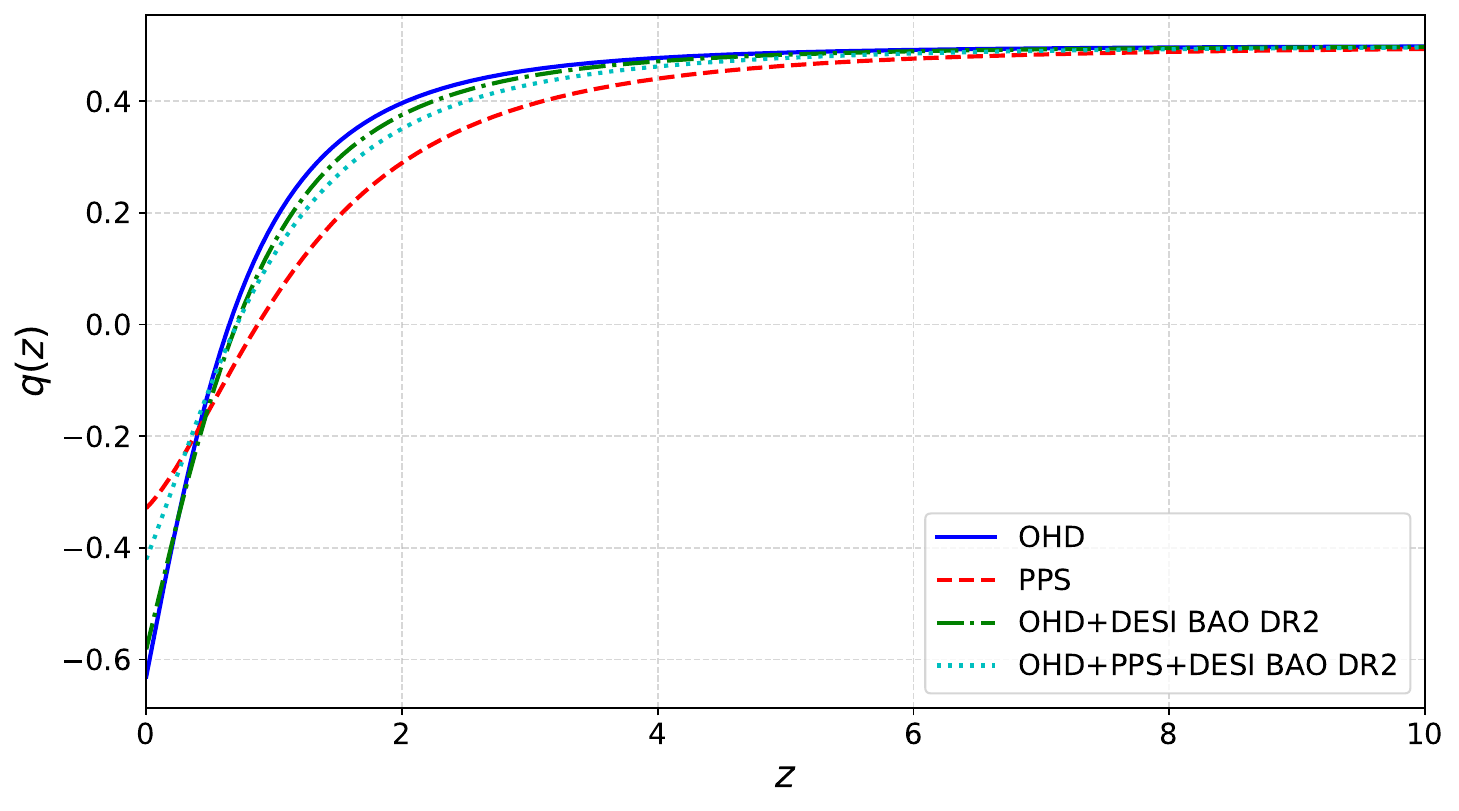}}
	\caption{Variation of deceleration parameter q versus redshift z.} 
\label{deceleration1}
\end{figure}

\begin{figure} [ht]
\centerline 
{ \includegraphics[scale=0.5]{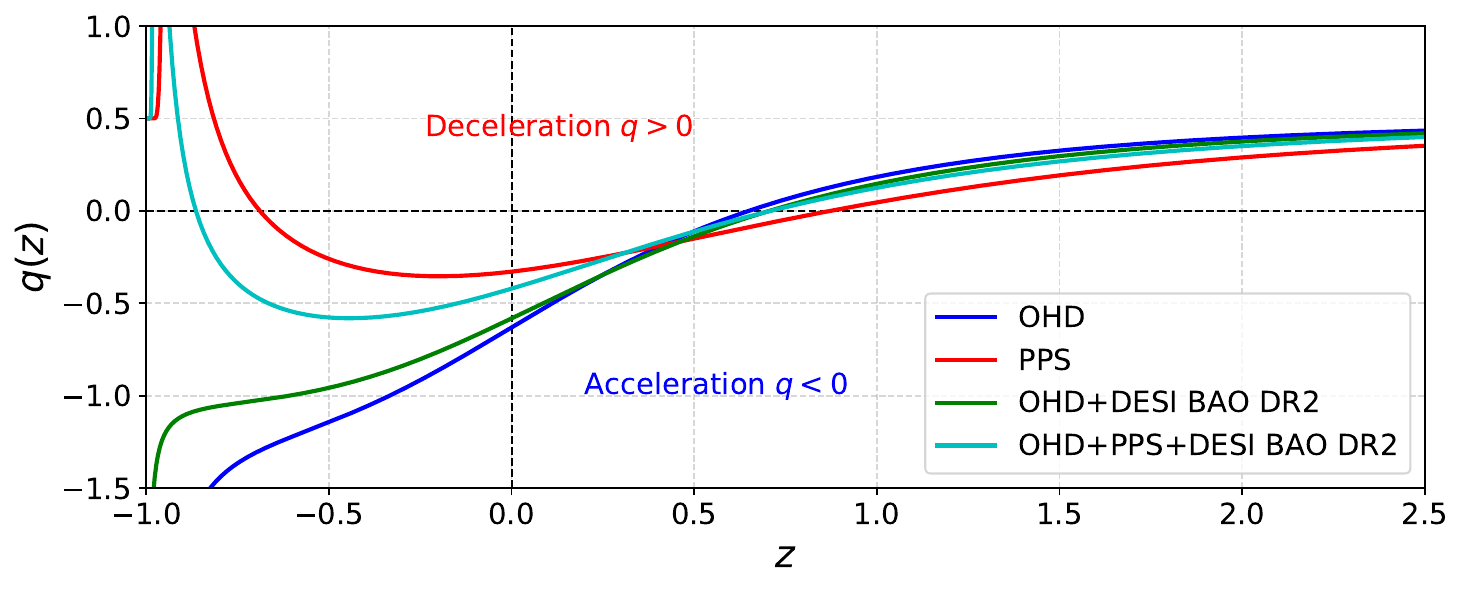}}
	\caption{{Redshift evolution of the deceleration parameter $q(z)$in the range $z \in \{-1, \, 2.5 \}$.}} 
\label{deceleration2}
\end{figure}

The evolution of the deceleration parameter as a function of redshift $z$ is shown in Fig.~\ref{deceleration1} for $z \in [0,  10]$ from the data combinations. Furthermore, we have also plotted the parameter $q$ in the range  $z \in [-1,2.5]$ in Fig.~\ref{deceleration2}, where we observed a second transition from acceleration to deceleration phase in the future where PPS data is included, namely for the PPS and OHD+PPS+DESI BAO DR2 combinations. This future transition disappeared in the case of the OHD and OHD+DESI BAO DR2 datasets, suggesting that this is not a generic feature of the model and is driven by the late-time luminosity distance constraints provided by the supernova dataset. The authors in~\cite{singh2024new} also find a second transition from acceleration to deceleration in the near future in all cases because of the OHD datasets used in the extended redshift range $z \sim 2.35$. In our analysis, we used the OHD points only from cosmic chronometers, and the future return to deceleration scenario disappeared. This supports that future deceleration is data data-driven feature, not a generic feature of the model. It is noteworthy that our model predicts a second future transition from acceleration back to deceleration occurring at a redshift $z_{\rm tr2}$, only when PPS data is employed, indicating that the supernova-based late-time observational data influences the inferred long-term future behavior of the cosmic expansion.
From an observational perspective, current data sets only constrain the expansion history up to $z\sim2$ and are not yet sensitive enough to strongly confirm or exclude such future evolution. Physically, a transition from acceleration to deceleration phase would imply that the universe may evolve toward a slower expansion phase in the distant future.

\begin{table}[ht]
\tbl{Summary of $q_0$, $z_{tr1}$, $z_{tr2}$, and $ t_0$ from different datasets.\label{tabparm2-dec}}
{\begin{tabular}{@{}lcccc@{}} \toprule
Dataset &  $ q_0$ & $ z_{tr1} $ & $z_{tr2}$ & $t_0$ \\ \colrule
OHD   &    $ -0.6305 $  &   $ 0.6494 $   &   $ - $ & $ 13.67 \pm 0.28$\\
PPS  &  $ -0.3289 $ &  $ 0.8721$  &  $ -0.6891$ &  $ 13.04 \pm 0.16$\\
OHD+DESI BAO DR2 &  $ -0.5818 $ &  $ 0.7053$  &  $ - $ &  $ 13.08 \pm 0.16$\\
OHD+PPS+DESI BAO DR2 &  $ -0.4207 $  &  $ 0.7087 $ &  $ -0.8639 $ &  $ 12.80 \pm 0.04$\\ 
\botrule
\end{tabular}}
\end{table}

The observational result of parameter value $q_0$ and transition redshifts are displayed in Table~\ref{tabparm2-dec}. The current value of the deceleration parameter confirms that the universe is undergoing accelerated expansion, with reported values ranging from $q_0 \approx -0.33$ to $q_0 \approx -0.63$, across different data combinations. When compared to recent literature, our results on $q_0$ from joint analysis are consistent with $q_0=-0.49$ by Luongo et al.~\cite{luongo2024model}, $q_0=-0.41$ by Iqbal et al.~\cite{iqbal2025exploring}, 
$q_0=-0.44$ by Bhagat et al.~\cite{bhagat2025exploring}.
The transition from deceleration to acceleration occurs at different redshifts from $z_{\rm tr1} \approx 0.65$ to $z_{\rm tr1} \approx 0.87$, depending on the datasets. 

\subsection{\textbf{\textit{Age of the universe}}}
The age of the universe in the considered model can be estimated by
\begin{equation}\label{age-1}
H_0 (t_0-t) = \int_{0}^{z}\frac{dz}{(1+z) E(z)}, \quad E(z) =\frac{H(z)}{H_0}.
\end{equation}
where $t_0$ is the present age of the universe, which can be obtained as
\begin{equation}
\label{age-2}
t_0 =\frac{1}{H_0} ~ \lim_{z\rightarrow\infty}\int_{0}^{z}\frac{dz}{(1+z) E(z)}.
\end{equation}

\begin{figure} [ht]
 \centerline{ \includegraphics[scale=0.5]{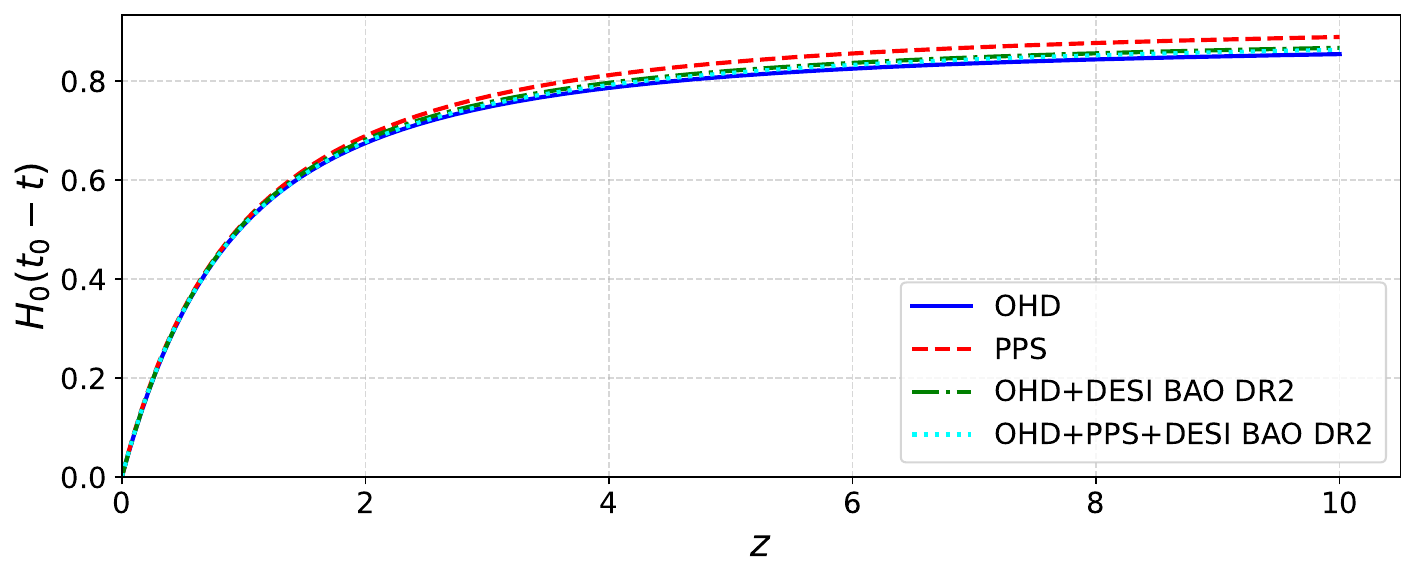}}
	\caption{{Variation of $H_0 (t_0-t)$ with redshift z.}} 
\label{age}
\end{figure}

The Fig.~\ref{age} represents the plot of redshift $z$ versus $H_0(t_0-t)$ from the datasets used. We have obtained the values of $H_0t_0 \approx 0.9$ for large values of redshift $z$. Using the best-fit values of $H_0$ from Table~\ref{params}, the present age can be estimated as $t_0 \approx 0.9(1/H_0)$. The estimated age of the universe for the derived model is mentioned in Table~\ref{tabparm2-dec}, where we observed that the mean value lies in the range from $t_0 \approx 12.80$ to $t_0 \approx  13.67$ in Gyrs. The findings on the present age of the universe are consistent with $t_0 = 13.28 \pm 0.33$ Gyrs by Singla et al.~\cite{singla2021accelerating}, $t_0 = 13.79 $ Gyrs by Goswami et al.~\cite{goswami2021modeling}, $t_0 = 13.65 $ Gyrs by Prasad et al.~\cite{prasad2020constraining} and Planck 2018 predictions, which yield $t_0 = 13.78 \pm 0.20$ Gyrs. For general and useful insights on the age of the universe, we refer the reader to see~\cite{DiValentino:2020srs}.

\subsection{\textbf{\textit{Om diagnose for null test}}}
\begin{figure} [ht]
 \centerline{\includegraphics[scale=0.5]{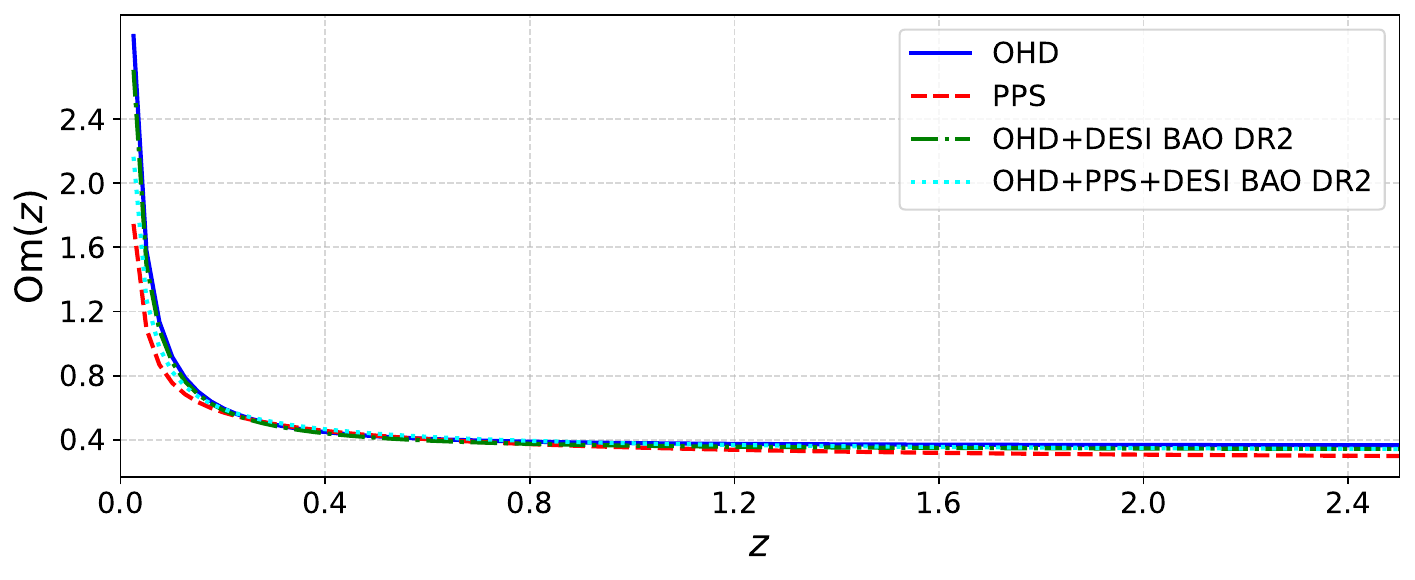}}
	\caption{{ Redshift evolution of the $Om(z)$ parameter.
    }} 
\label{om}
\end{figure}

The $Om(z)$ parameter in the $\Lambda$CDM model is a straight line plotted against $z$ and is constant at all redshifts.  Any deviation from this straight line indicates the existence of a dynamical DE component. In particular, a curve above the $\Lambda$CDM line indicates a quintessence-like behavior ($w>-1$), while a curve below indicates a phantom behavior ($w<-1$). The $Om(z)$ parameter~\cite{sahni2008two, sahni2014model} is defined as
\begin{equation}
\label{Om}
Om(z) = \frac{E^2(z)-1}{(z+1)^3-1}.
\end{equation}
The Fig.~\ref{om} indicates the redshift evolution of $Om(z)$ based on observational data sets. It can be seen from Fig. \ref{om} that $Om(z)$ is positive for all redshifts $z \in [0,2.5]$ and increases with decreasing redshift. This contrasts with the $\Lambda$CDM prediction, where the $Om(z)$ parameter is constant. The increasing nature of $Om(z)$ in our analysis may suggest the presence of a dynamical DE component. Furthermore, the positive value of $Om(z)$ at $z=0$ differs from the $\Lambda$CDM prediction, where the parameter approaches a null value~\cite{shahalam2015om}.

\section{Conclusion} \label{model_conclusion}

In this work, we have investigated a new parametrization of the dark energy equation of state, which approaches the $\Lambda$CDM model for parameter value, $\alpha=0$. We have derived precision constraints on the model parameters using the recent observational datasets:  OHD, PPS, and in combination, OHD + DESI BAO DR2, OHD + PPS + DESI BAO DR2. 
We have obtained the best fit value of parameter, $\alpha=0.239 \pm 0.07$ at 68\% CL from joint analysis, which is non-null and suggests deviation from $\Lambda$CDM cosmology. The model accommodates varying values of the Hubble constant from different datasets and joint analysis yields $H_0 = 68.40 \pm 0.23\, \mathrm{Km\,s^ {-1} Mpc^{-1}}$ at 68\% CL, which is higher than the Planck measurement estimate~\cite{Planck:2018vyg}. The derived constraints on the Hubble constant are more robust and precise than those reported in~\cite{singh2024new} and also consistent with recent findings in the literature, as mentioned in Section. \ref{results}. We examined the physical behavior of the model by analyzing the deceleration parameter, the age of the universe, and the $Om(z)$ diagnose. The deceleration parameter confirms a smooth transition from the past deceleration phase to the present cosmic acceleration and also a second future transition back to deceleration only when PPS data is employed. The present value of deceleration parameters is obtained as $q_0=-0.4207$ from joint analysis, see Table \ref{tabparm2-dec} and Fig. \ref{deceleration2}. The age of the universe in the model is estimated as $t_0 \approx 13$ Gyrs, consistent with the Planck 2018 prediction~\cite{Planck:2018vyg}. The analysis of the $Om(z)$ parameter shows that the universe in the investigated model has a marginal departure from the $\Lambda$CDM universe. As a final remark, we observe that the investigated model shows a minimal departure from the $\Lambda$CDM universe in the case of joint analysis and provides robust estimates of $H_0, q_0$, and $t_0$, as compared to the findings in~\cite{singh2024new}, in the light of recent data used. The future studies incorporating additional observational probes, including  CMB, SPT, and ACT data, could help to further constrain the model parameters and evaluate its compatibility with a wider spectrum of cosmological observations.

\bibliographystyle{ws-ijgmmp}
\bibliography{refs}

\end{document}